\documentclass{ifacconf}
\usepackage{pdf14}
\usepackage{cite}
\usepackage{amsmath,amssymb,amsfonts}
\usepackage{algorithmic}
\usepackage{textcomp}
\usepackage[printonlyused,nolist]{acronym}
\usepackage[inline]{enumitem}
\usepackage{xcolor}
\usepackage{graphicx}      
\usepackage{natbib}        

\begin{document}
\begin{frontmatter}

\title{
Towards more realistic co-simulation of cyber-physical energy distribution systems
}

\author[First,Second]{Immanuel Hacker}
\author[First,Second]{Ömer Sen}
\author[Second]{Dennis van der Velde}
\author[First,Second]{Florian Schmidtke}
\author[First,Second]{Andreas Ulbig}

\address[First]{IAEW at RWTH Aachen University, Aachen, Germany\\(e-mail: \{i.hacker, o.sen, f.schmidtke, a.ulbig\}@iaew.rwth-aachen.de).}
\address[Second]{Fraunhofer FIT, Aachen, Germany\\(e-mail: dennis.van.der.velde@fit.fraunhofer.de)}


\begin{abstract}                
The increased integration of information and communications technology at the distribution grid level offers broader opportunities for active operational management concepts.
At the same time, requirements for resilience against internal and external threats to the power supply, such as outages or cyberattacks, are increasing.
The emerging threat landscape needs to be investigated to ensure the security of supply of future distribution grids.
This extended abstract presents a co-simulation environment to study communication infrastructures for the resilient operation of distribution grids.
For this purpose, a communication network emulation and a power grid simulation are combined in a common modular environment.
This will provide the basis for cybersecurity investigations and testing of new active operation management concepts for smart grids.
Exemplary laboratory tests and attack replications will be used to demonstrate the diverse use cases of our co-simulation approach.
\end{abstract}


\end{frontmatter}

\begin{acronym}
    \acro{BMWK}{German Federal Ministry for Economic Affairs and Climate Action}
    \acro{DSO}{Distribution System Operator}
    \acro{ICT}{Information and Communications Technology}
    \acro{IDS}{Intrusion Detection System}
    \acro{IT}{Information Technology}
    \acro{OT}{Operational Technology}
    \acro{ET}{Energy Technology}
    \acro{EMS}{Energy Management System}
    \acro{RTU}{Remote Terminal Unit}
    \acro{SCADA}{Supervisory Control and Data Acquisition}
    \acro{SMGW}{Smart Meter Gateway}
    \acro{TSO}{Transmission System Operator}
    \acro{SGAM}{Smart Grid Architecture Model}
    \acro{PCAP}{Packet Capture}
    \acro{vRTU}{virtual RTU}
    \acro{MTU}{Master Terminal Unit}
    \acro{DER}{Decentralized Energy Resource}
    \acro{VED}{Virtual Edge Devices}
    \acro{DMZ}{Demilitarized Zone}
    \acro{IED}{Intelligent Electronic Device}
    \acro{VPN}{Virtual Private Network}
    \acro{VPP}{Virtual Power Plant}
    \acro{VED}{Virtual Edge Device}
    \acro{SG}{Smart Grid}
    \acro{CVE}{Common Vulnerabilities and Exposures}
    \acro{RCE}{Remote Code Execution}
    \acro{PE}{Privilege Escalation}
    \acro{SAM}{Simulated Attacker Model}
    \acro{SUID}{set-user-ID}
    \acro{FDI}{False Data Injection}
    \acro{SG}{Smart Grid}
\end{acronym}



\section{Introduction} \label{sec:introduction}

In distribution grids, progressive digitization is leading to increasing integration of \ac{ICT}.
This provides enhanced grid transparency through more opportunities for real-time monitoring of the grid and also offers opportunities for active control of individual assets of the overall system.
This opens up new potential in the area of active operational management at the distribution grid level.\newline
In particular, flexibility resources provided by new types of (distributed) generation and consumers at this grid level should be used in an optimizing manner, both market- and grid-related.
Multi-use concepts developed for this purpose are the subject of current research~\citep{truong2018multi}.\newline
In addition to the acquisition of new potentials, the increasing inclusion of \ac{ICT} in grid-relevant operations also requires the fulfillment of fundamental requirements for cyber-resilience and cybersecurity of the overall system.
Malfunctions in the \ac{ICT} domain can have a direct impact on the security of supply.
In particular, adequate measures are needed to prevent cyberattacks to counter the increased attack surface.
This requires providing communications data from attacks patterns that can be used to develop, validate, and test such measures.
However, such data is not publicly available~\citep{3_zuech2015intrusion}.

Both the development of new operational management concepts and the investigation of domain-specific cybersecurity measures require tailored development environments that share essential requirements in terms of their scalability, realism (e.g.,\ digital twin), and flexibility.

This paper presents our ongoing work on a framework for the co-simulation of cyber-physical energy systems.
For this, we present the goals and added values of the environment in Section~\ref{sec:Background}.
Section~\ref{sec:cosim} gives an overview of the environment and its main components.
The focus of the paper is on the presentation of two exemplary use cases in Section~\ref{sec:usecase}, which we implement with the help of the environment.
This includes the replication of cyberattacks to provide a development environment for countermeasures, e.g.,\ for intrusion detection, and the use of the environment as a digital twin for practical testing of operational concepts for flexibility management.


\section{Background \& Related Work} \label{sec:Background}

\begin{figure*}
    \centering
    \includegraphics[width=\linewidth]{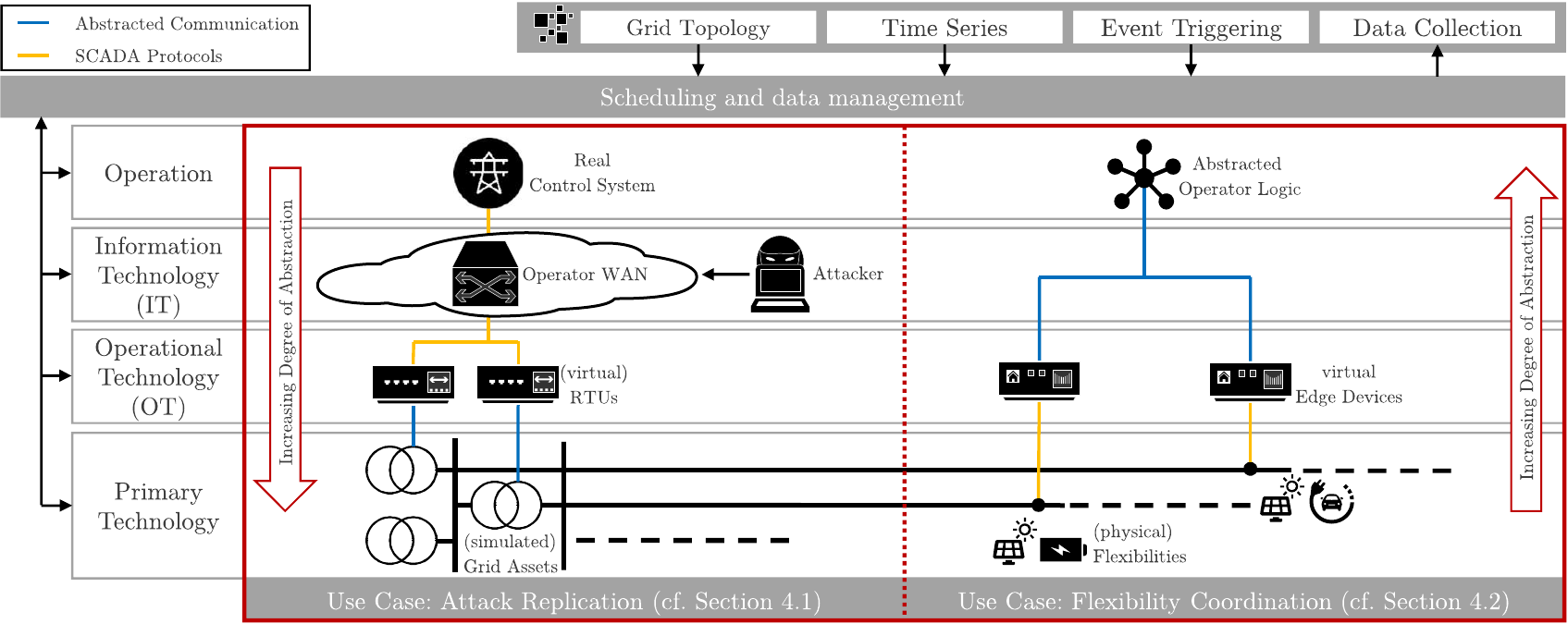}
    \caption{Overview of the utilized co-simulation environment and its major components.
    Depending on the use case, the respective components can be modeled in different degrees of abstraction.}
    \label{fig:cosim}
\end{figure*}

In this section, we introduce the important parts of a \ac{SG} infrastructure to be considered in a simulation environment for our use case and derive the requirements.
First, we give a brief overview of \ac{SG}s and their simulation, and then their cybersecurity.\newline
Usually, the \ac{SG} infrastructure is divided into different zones.
Each component of the energy system refers to one of those zones.
The important zones for our work are Primary Technology, \ac{OT}, \ac{IT} and Operation.
Primary Technologies are power grid equipment, such as transformers, switches, loads, and generation units.
Those assets are connected to the \ac{OT} network to control them or to retrieve information/measurements.
The \ac{IT} section represents the communication infrastructure used by the \ac{OT} components for the communication with operators, such as a \ac{DSO}.\newline
A simulation environment needs to meet specific requirements for the appropriate modelling of the \ac{SG} infrastructure proposed before.
First, mutual dependencies between the zones need to be considered.
Second, we need a modular, component-based infrastructure model.
Hereby, separate models with different detail level should be considered.

There are already various approaches in current research that depict \ac{SG}s both using hardware within laboratories and software in the sense of a simulation.
~\cite{cosim_sil_2020} utilize a Software-in-the-Loop co-simulation for the preliminary testing of \ac{SG} software rollouts.
In~\cite{astoria_2016} a co-simulation is used to simulate cyberattacks in \ac{SG}s.
As part of our developments, we needed a common environment in which we could flexibly connect both hardware and software components.
Only in this manner can we investigate all use cases and interactions relevant to us.\newline
Despite a variety of preventive measures such as strict user management, password policies, access control, and network segmentation, intrusion detection capabilities are still needed to meet the high-security requirements in \ac{SG}s.
A holistic security concept, consisting not only of preventive measures but also of detective and reactive measures, can provide options for counteracting intruders even in cases where they are not prevented.
Several approaches address the analysis of cybersecurity by investigating cyberattack scenarios, e.g., denial-of-service attacks, false data injection, or a physical system disruption technique, using mathematical modeling or cyberattack trees~\citep{22_falco2018master}.
In addition, several co-simulation approaches are being explored that consider hardware-in-the-loop co-simulation or a synthetic framework to simulate attack scenarios in \ac{SG} applications to generate normal and attack data~\citep{23_albarakati2018openstack}.
Our approach leverages isolated and real-time co-simulation so that we can perform a variety of services and vulnerabilities that allow us to simulate a dynamic, multi-stage attacker exploring the network and exploiting found vulnerabilities in a virtual, as well as a cyber-physical, environment.


\section{Co-Simulation for Smart Grids} \label{sec:cosim}

This Section introduces the overall environment we use to model the relevant use cases for smart distribution grids.
Section~\ref{subsec:cosim_overview} provides an overview of this environment.
In Section~\ref{subsec:cosim_components} we describe the environment components.

\subsection{Environment Overview} \label{subsec:cosim_overview}
Our work is based on a co-simulation environment and a \ac{SG} laboratory~\cite{van2021towards}.
The framework \textit{mosaik} is used as a modelling tool~\citep{mosaik_2019}.
It allows the connection of various independent simulators.
The \textit{mosaik} framework schedules the time-discrete proceeding of each simulator and manages the data flow of connecting simulators.
Therefore, a dynamic interconnection of those simulators is possible.
In this way, the data flow as abstracted communication is considered explicitly.
In addition, we can connect real components of our \ac{SG} laboratory to the environment.
Therefore, it is possible to use both simulated and real components.
Although, parts of the laboratory can be combined with simulated parts.
Figure~\ref{fig:cosim} shows a general overview of the co-simulation environment.

As mentioned before, the different zones Primary Technology, \ac{OT}, \ac{IT} and Operation are considered in our work.
The depth of implementation of the components in each zone can vary depending on the use case.
In our work, there are two aspects that we primarily focus on.
On the one hand, we focus on attack replications.
For this use case, the zones at the lower level are more abstracted while operation processes and \ac{IT} are modelled in detail.
On the other hand, we focus on flexibility coordination processes for small-scale \ac{DER}.
Therefore, we implement a bottom-up modelling approach for flexibility units and operation technologies at the household level.\newline
In addition to those use case-specific implementation, we use general implementations for all use cases.
These implementations are axillary functions for the preparation of time series data and grid topologies, data collection, and a power flow simulation based on \textit{PandaPower}~\citep{pandapower_2018}.

\subsection{Components and Functionality} \label{subsec:cosim_components}
In the following, the functionalities of the most important components of the deployed environment for the use cases discussed in the paper are explained in more detail.\newline
To map the \ac{IT} and \ac{OT} networks and their components such as switches, routers or firewalls, we use the emulation environment Containernet~\citep{containernet_2016}.
The nodes of the communication networks are containerized here and enable flexible deployment of the various components.
Deployment and synchronization with the overall environment is done via an interface with \textit{mosaik}.\newline
\ac{vRTU}s can also be deployed within the nodes.
These enable communication with real \ac{SCADA} protocols (IEC~60870-5-104).
The data points of the protocol can be logically linked and additional services (SSH, SNMP, \ldots) can be added.
These services can also be specifically equipped with vulnerabilities that serve as initial access for the replication of cyberattacks.\newline
We use the concept of \ac{VED} for modelling Smart Homes.
Using this concept, we explicitly consider well-defined interfaces of Smart Homes for the communication with external actors as well as for the control of behind-the-meter assets like Battery Storage and Photovoltaic systems.
An Energy Management System bundles retrieved data implements the operation logic and sends new control commands for connected assets.
Within the lab, the communication with inverters is via Modbus.\newline
The option of using real-world communication protocols both between \ac{OT}-layer and operational layer and between \ac{OT}-layer and primary technology allows flexible use of either simulated or real components on each of the layers.
This allows us to use components from our \ac{SG} Laboratory on all layers if required.
This includes a control system, several \acp{RTU} and measuring systems concerning the \acp{ICT} and a flexible distribution grid with medium/low-voltage substations and flexibilities such as controllable loads, storage systems, and photovoltaic plants.


\section{Use Cases and Framework Integration} \label{sec:usecase}

Based on the environment previously introduced, various investigations can be performed.
In this paper we focus on the two use cases replication of cyberattacks (cf. Section~\ref{subsec:usecase_impact}) and the development and testing of operational management concepts for grid flexibilities (cf. Section~\ref{subsec:usecase_operation}).

\subsection{Cyberattack Replication in Smart Grids} \label{subsec:usecase_impact}
\begin{figure}[t]
    \centerline{\includegraphics[width=\columnwidth]{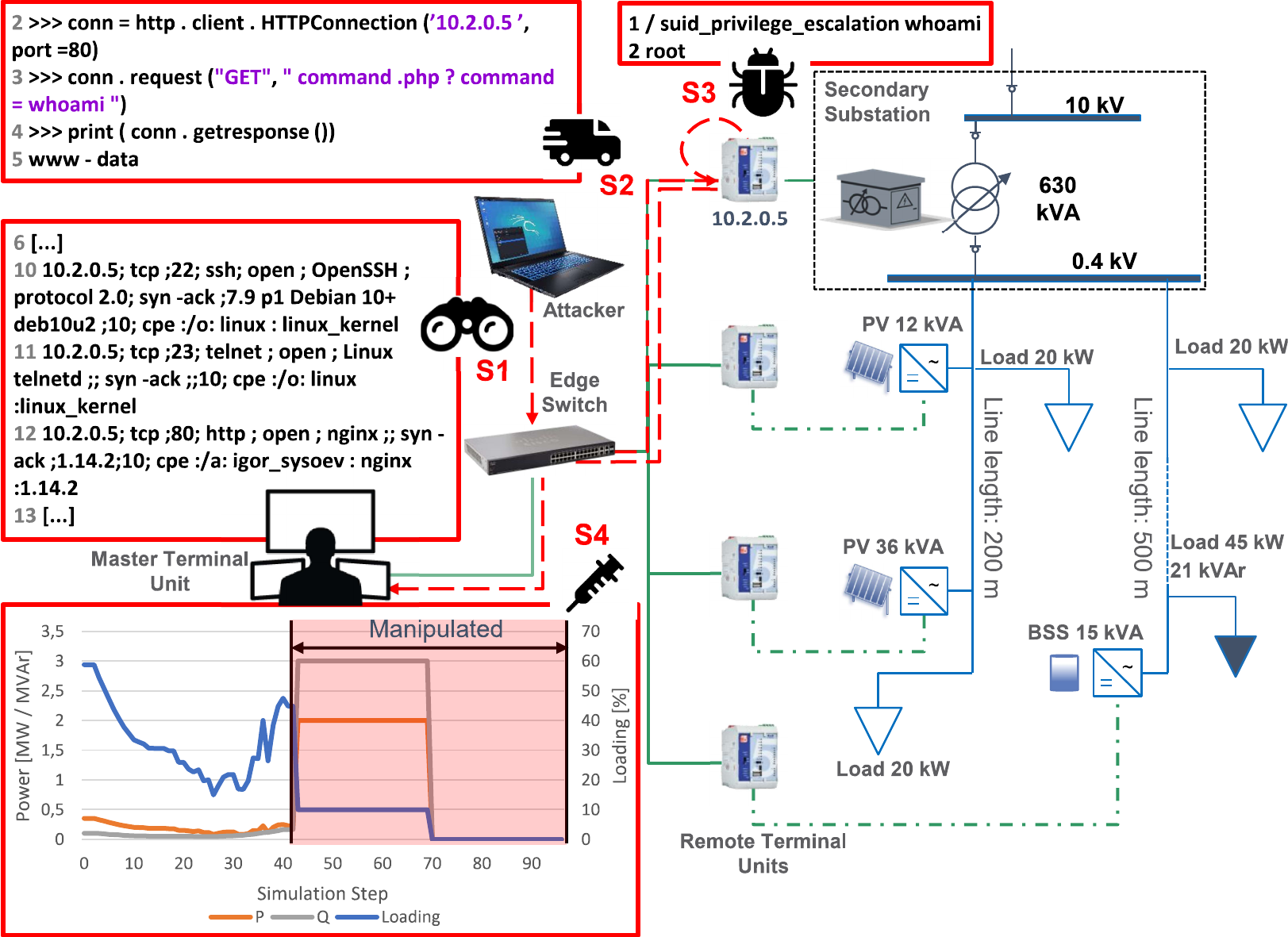}}
    \caption{Illustration of the simulated attack replication scenario performing an exemplary multi-stage attack~\citep{sen2021approach}:
    network scan (S1), \ac{RCE} (S2), \ac{PE} (S3), and data manipulation (S4).
    }
    \label{fig:result_plot_attack}
\end{figure}

Based on our previous work~\citep{sen2021approach}, an example multi-stage attack scenario is demonstrated using a simulated medium/low voltage distribution grid equipped with networked assets such as edge switch, \ac{MTU}, \ac{RTU}, \ac{DER}, etc. (cf.~Figure~\ref{fig:result_plot_attack}).
Here, the \ac{RTU} components are equipped with vulnerabilities that allow non-legitimate remote control via attack vectors such as \ac{RCE} and \ac{PE}.
The concept of these vulnerabilities is based on providing simple running network services such as a web interface, SSH server, or Telnet service that allow executing commands on the computer system.
After gaining remote access via \ac{RCE} vulnerabilities, the \ac{SAM} then attempts to escalate its privileges on the infiltrated, e.g., Linux-based, host systems by exploiting \ac{PE} vulnerabilities such as \ac{SUID} or a sudoers-enabled script that allows the execution of administrative commands without authentication.

For further illustration, the automatically executed phases of the attack process within the co-simulation are shown using terminal executions and the effects of the attack are shown by plotting the measurements across the simulation steps.
In the first phase of the sequence, the \ac{SAM} scans the \ac{SCADA} network and identifies the connected device, e.g., a \ac{RTU} on a secondary substation transformer with its open TCP ports indicating configured services such as SSH, Telnet, and Nginx web server.
Based on the collected information, in the next phase, \ac{SAM} executes a \ac{RCE} over the identified vulnerable web interface on port 80 equipped with a command execution script by specifying the command 'whoami' as a parameter, which issues the executing user 'www-data'.
Since the user 'www-data' has no administrative privileges, the \ac{SAM} extends its privileges through \ac{PE} by exploiting a found \ac{SUID} vulnerability.
In the final phase, the \ac{SAM} manipulates the measured load, active and reactive power data from the secondary substation and transmits it to the~\ac{MTU}.
This demonstration illustrates the capabilities of \ac{SAM} to manipulate process data remotely.
With intelligent manipulation strategies based on \ac{FDI} techniques~\citep{26_sayghe2020survey}, \ac{SAM} could even disrupt grid operations over the long term.

\subsection{Testing of Operational Management Concepts} \label{subsec:usecase_operation}
One of the biggest challenges of future active distribution grids is coordinating behind-the-meter flexibility.
This is all flexibility on the customer premise like heat pumps, electric cars, and battery storage systems.
Our work on operational flexibility mainly focuses on multi-use strategies and the cyber-resilience of these connected systems.
Different stakeholders are interested in using this operational flexibility,  mainly the owner of the asset who wants to use the primary purpose with minimal costs, the \ac{DSO} who wants to optimize the grid operation, and market actors like \ac{VPP} who want to monetize the flexibility.
The goal of multi-use strategies is to find an optimal solution to balance the interests of the different stakeholders.
These systems require a high degree of networking, so cyber-resilience is essential.
\ac{EMS} are at the core of this concept, they implement the optimization logic and take over the control of the assets and the communication with external platforms.
In~\citep{hacker2021framework} we showed how the development of \ac{EMS} logic could take place in a digital twin and then be transferred to our laboratory grid.
\begin{figure}[t]
	\centerline{\includegraphics[width=\columnwidth]{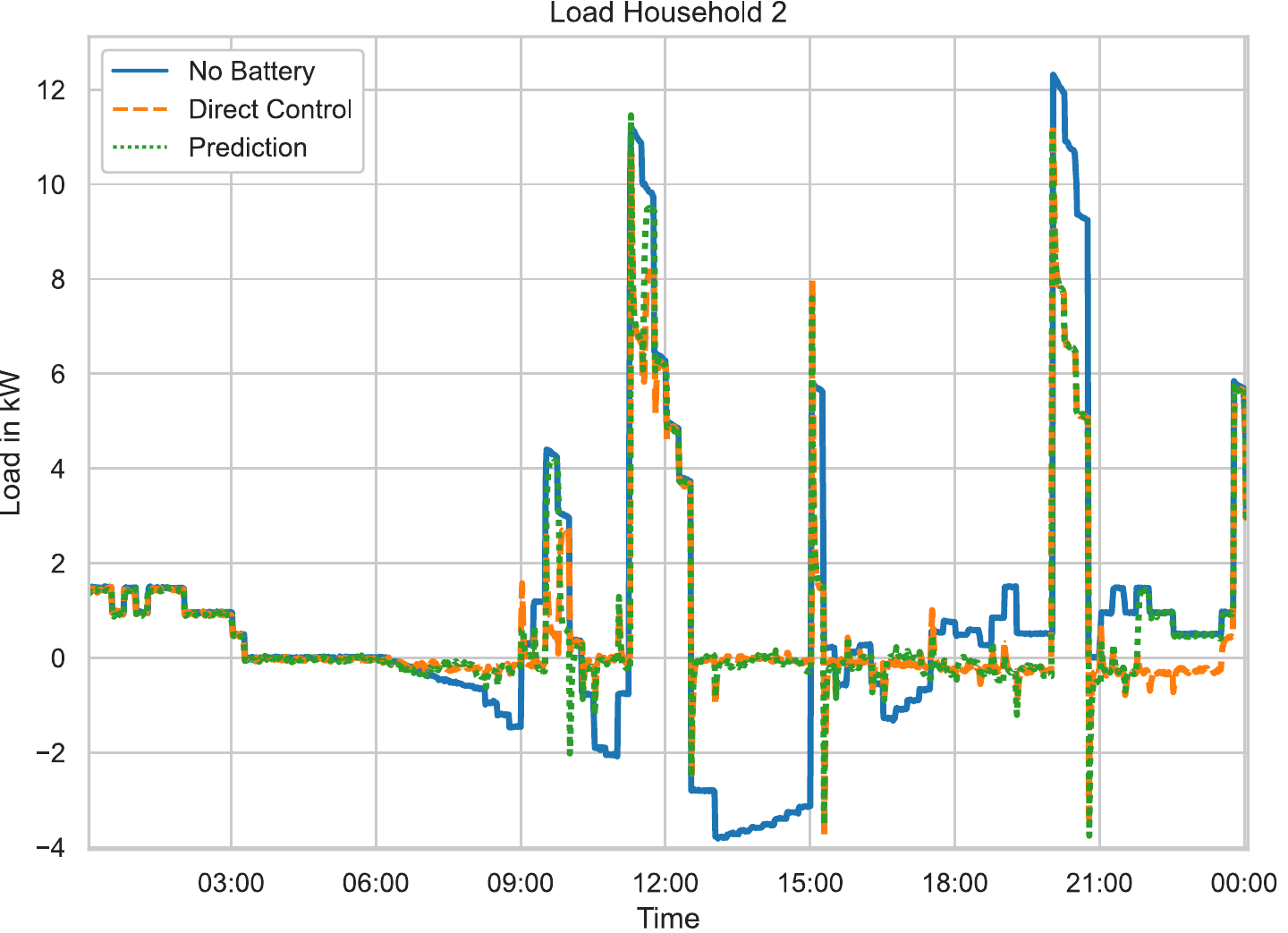}}
	\caption{Exemplary results of operational management strategies in the \ac{SG} Laboratory.\citep{hacker2021framework}
	}
	\label{fig:result_plot_operational_management}
\end{figure}

Figure~\ref{fig:result_plot_operational_management} shows the results of using different logic in the \ac{SG} Laboratory integrated by the \ac{VED} (cf. Section~\ref{subsec:cosim_components}).
In addition to the modular exchange of the connection with the assets, coordination platforms can also be abstracted, or prototypes of real platforms can be used.
This allows the future analysis of the cyber-resilience of different \ac{IT}-Architectures.

\section{Conclusion} \label{sec:conclusion}
In this paper, we present our co-simulation environment used for the investigation of domain-specific use cases such as cybersecurity and \ac{SG} operation concepts.
Our proposed modular approach is used for simulative evaluation and evaluation in our \ac{SG} Laboratory.
This allows adjusting the level of abstraction of all components depending on the use cases.
We have shown that the environment can be used productively for both of the use cases presented.
Regarding attack replication, we have shown how we can perform manipulations of data traffic.
In future works, we can use this to generate attack data that can be used to develop countermeasures such as intrusion detection.
In addition, we have also shown how we can integrate operational management concepts for controlling decentralized flexibilities into the environment.
For us, this forms the basis for further developments in the area of multi-use strategies and the cyber-resilience of such networked systems.

\begin{ack}
\vspace{-1em}
\begin{minipage}{0.65\columnwidth}%
This work has received funding from the \ac{BMWK} under project funding reference 0350056 (FlexHub).
\end{minipage}
\hspace{0.02\columnwidth}
\begin{minipage}{0.35\columnwidth}%
	\includegraphics[width=\textwidth]{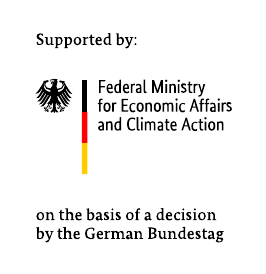}
\end{minipage}
\end{ack}

\vspace{-1em}
\bibliography{ifacconf}

\begin{thebibliography}{13}
\providecommand{\natexlab}[1]{#1}
\providecommand{\url}[1]{\texttt{#1}}
\providecommand{\urlprefix}{URL }
\expandafter\ifx\csname urlstyle\endcsname\relax
  \providecommand{\doi}[1]{doi:\discretionary{}{}{}#1}\else
  \providecommand{\doi}{doi:\discretionary{}{}{}\begingroup
  \urlstyle{rm}\Url}\fi

\bibitem[{{Albarakati, Abdullah et al.}(2018)}]{23_albarakati2018openstack}
{Albarakati, Abdullah et al.} (2018).
\newblock Openstack-based evaluation framework for smart grid cyber security.
\newblock In \emph{IEEE SmartGridComm}.

\bibitem[{{Falco, Gregory et al.}(2018)}]{22_falco2018master}
{Falco, Gregory et al.} (2018).
\newblock A master attack methodology for an ai-based automated attack planner
  for smart cities.
\newblock In \emph{IEEE Access}.

\bibitem[{{Hacker, Immanuel et al.}(2021)}]{hacker2021framework}
{Hacker, Immanuel et al.} (2021).
\newblock A framework to evaluate multi-use flexibility concepts simultaneously
  in a co-simulation environment and a cyber-physical laboratory.
\newblock In \emph{CIRED 2021}.

\bibitem[{{Peuster, M. et al.}(2016)}]{containernet_2016}
{Peuster, M. et al.} (2016).
\newblock {MeDICINE: Rapid Prototyping of Production-Ready Network Services in
  Multi-PoP Environments}.
\newblock In \emph{NFV-SDN}.

\bibitem[{{Sayghe, Ali et al.}(2020)}]{26_sayghe2020survey}
{Sayghe, Ali et al.} (2020).
\newblock Survey of machine learning methods for detecting false data injection
  attacks in power systems.
\newblock In \emph{IET Smart Grid}.

\bibitem[{{Sen, {\"O}mer et al.}(2021)}]{sen2021approach}
{Sen, {\"O}mer et al.} (2021).
\newblock An approach of replicating multi-staged cyber-attacks and
  countermeasures in a smart grid co-simulation environment.
\newblock In \emph{CIRED 2021}.

\bibitem[{{Steinbrink, Cornelius et al.}(2019)}]{mosaik_2019}
{Steinbrink, Cornelius et al.} (2019).
\newblock {CPES Testing with mosaik: Co-Simulation Planning, Execution and
  Analysis}.
\newblock \emph{Applied Sciences}.

\bibitem[{{Thurner, L. et al.}(2018)}]{pandapower_2018}
{Thurner, L. et al.} (2018).
\newblock {pandapower — An Open-Source Python Tool for Convenient Modeling,
  Analysis, and Optimization of Electric Power Systems}.
\newblock \emph{IEEE Transactions on Power Systems}.

\bibitem[{{Truong, Cong Nam et al.}(2018)}]{truong2018multi}
{Truong, Cong Nam et al.} (2018).
\newblock Multi-use of stationary battery storage systems with blockchain based
  markets.
\newblock \emph{Energy Procedia}.

\bibitem[{{van der Velde, Dennis et al.}(2021)}]{van2021towards}
{van der Velde, Dennis et al.} (2021).
\newblock Towards a scalable and flexible smart grid co-simulation environment
  to investigate communication infrastructures for resilient distribution grid
  operation.
\newblock In \emph{SEST}.

\bibitem[{{Veith, Eric et al.}(2020)}]{cosim_sil_2020}
{Veith, Eric et al.} (2020).
\newblock {Large-Scale Co-Simulation of Power Grid and Communication Network
  Models with Software in the Loop}.
\newblock In \emph{ENERGY}.

\bibitem[{{Wermann, A. G. et al.}(2016)}]{astoria_2016}
{Wermann, A. G. et al.} (2016).
\newblock {ASTORIA: A Framework for Attack Simulation and Evaluation in Smart
  Grids}.
\newblock In \emph{NOMS - IEEE/IFIP}.

\bibitem[{{Zuech, Richard et al.}(2015)}]{3_zuech2015intrusion}
{Zuech, Richard et al.} (2015).
\newblock Intrusion detection and big heterogeneous data: a survey.
\newblock In \emph{Journal of Big Data}.

\end{thebibliography}

\end{document}